# Superconductivity in Shear Strained Semiconductors

Chang Liu[1,2], Xianqi Song[1], Quan Li[1,2,*], Yanming Ma[1,2,*] and Changfeng Chen[3,*]

[1]State Key Laboratory of Superhard Materials, and International Center for Computational Methods and Software, College of Physics, Jilin University, Changchun 130012, China.

[2]International Center of Future Science, Jilin University, Changchun 130012, China.

[3]Department of Physics and Astronomy, University of Nevada, Las Vegas, Nevada 89154, USA.

∗**Corresponding author.** email: liquan777@calypso.cn; chen@physics.unlv.edu; mym@jlu.edu.cn

ABSTRACT

Semiconductivity and superconductivity are remarkable quantum phenomena that have immense impact on science and technology, and materials that can be tuned, usually by pressure or doping, to host both types of quantum states are of great fundamental and practical significance. Here we show by first-principles calculations a distinct route for tuning semiconductors into superconductors by diverse large-range elastic shear strains, as demonstrated in exemplary cases of silicon and silicon carbide. Analysis of strain driven evolution of bonding structure, electronic states, lattice vibration, and electron-phonon coupling unveils robust pervading deformation induced mechanisms auspicious for modulating semiconducting and superconducting states under versatile material conditions. This finding opens vast untapped structural configurations for rational exploration of tunable emergence and transition of these intricate quantum phenomena in a broad range of materials.

**Keywords:** Semiconductors; Electron-phonon coupling; Strain modulation; First-principles calculations; Superconductivity

## Introduction

Semiconductors and superconductors exhibit profound and distinct quantum phenomena. Materials that can host both behaviors offer an appealing platform for innovative research and development [1]. The quest to explore superconducting states in semiconductors has a long history dating back to 1960s [2-7] Most works have taken two established routes: (i) external doping by atomic insertion or substitution or self-doping by alloying [8], and (ii) application of high pressure [9, 10]. The resulting electrically conducting states couple to another degree of freedom, e.g., lattice vibration (i.e., phonon), to generate superconductivity in chemically (doping) or physically (pressure) modulated semiconductors. In recent years, strain engineering has been employed to induce superconductivity at surfaces or thin films via adatom adsorption or interfacial lattice mismatch [11-15]. This approach is effective, but lacks flexibility of tuning the resulting superconducting states, since the strains are fixed once the surface or substrate conditions are set. It is desirable to explore further avenues for achieving more flexible and tunable modulation of semiconducting and superconducting states, preferably reversibly over large strain ranges and in ways that are compatible with nanoelectronic synthesis and characterization protocols.

Recent research revealed that diamond can be driven into a metallic state by biaxial compression-shear (CS) deformation, and charge redistribution and phonon softening at rising strains generate superconductivity in this ultralarge-bandgap insulator [16]. This surprising result suggests that shear-strain tuning may offer an effective tool for generating and tuning superconductivity in semiconductors. There are, however, crucial differences in structural and electronic properties distinguishing most semiconductors from diamond. In particular, diamond's unique tendency of graphitization accompanied by the associated large bonding changes plays a key role in inducing strong charge redistribution and phonon softening that drive superconductivity in its deformed crystal [16]. It is therefore compelling to probe how typical prominent semiconductors would deform under various shear strains and whether metallic and superconducting states would emerge, and if affirmative, how would the driving mechanism be related yet distinct compared to that in deformed diamond and, most crucially, whether these insights are generally applicable to wide-ranging semiconductors.

In this work, we report findings from first-principles calculations on emerging superconducting states in shear strained classic semiconductors Si and SiC. In contrast to diamond that develops superconductivity under special CS strains, Si and SiC host superconducting states under diverse shear conditions, including unconstrained pure shear strains, offering expanded tunability for flexible and versatile experimental control and implementation. We analyze evolution of structural, electronic, phonon, and electron-phonon coupling with changing strains in elastically deformed Si and SiC and identify robust metallization and phonon softening mechanisms that should be applicable to a broad variety of materials. These findings showcase elastic shear-strain engineering as an effective tool for tuning semiconducting and superconducting states over large strain ranges, opening new access to hitherto little explored deformed material configurations for discovery of fascinating physics phenomena.

Silicon is known to undergo a series of structural transitions from the cubic diamond phase to *β*-tin, simple hexagonal (sh), and then hexagonal close packed (hcp) phases at increasing pressures [17-19], and these high-pressure phases of Si are all metallic and superconducting with a peak $T_c$ of 8.2 K reached at 15 GPa in the sh-Si phase [20, 21]. The cubic diamond phase of Si also can host superconductivity under carrier doping via boron-atom insertion with $T_c$≈0.35 K [22], which is an order of magnitude lower than that observed in boron doped diamond [23-25] due to much weaker bonding strength and lower phonon frequency in Si. Superconductivity has been induced in the cubic diamond phase of SiC by carrier doping through boron-atom insertion with $T_c$≈1.4 K [26]. There has been no report on pressure-induced superconductivity in the wide-bandgap semiconductor SiC. In this work, we show that full-range elastic shear-strain tuning offers an alternative and versatile means to produce and modulate superconductivity in Si and SiC, and that the obtained results suggest common and robust mechanisms for wide-ranging semiconductors.

**Methods**

For stress-strain calculations, we have employed VASP code [27] based on the density functional theory (DFT) in the local density approximation (LDA) [28, 29]. We also have calculated lattice dynamics and electron-phonon coupling (EPC) using the density-functional perturbation theory (DFPT) in linear response as implemented in QUANTUM ESPRESSO code [30]. Such calculations produce adequate structural and mechanical properties and describe well metallic states and produce fairly accurate $T_c$ in metallized semiconductors, as extensively demonstrated in boron-doped diamond and silicon [23, 31-37]. LDA calculations, however, are known to underestimate electronic bandgaps in semiconductors, so we have adopted the Heyd-Scuseria-Ernzerhof (HSE) hybrid functional [38, 39] that accurately reproduces experimental bandgaps of Si and SiC [40, 41]. More computational details are given in in Supplementary Information. From Eliashberg theory of superconductivity [42, 43], McMillan derived [44], later modified by Allen and Dynes [45], an analytic expression for transition temperature,

$$T_c = \frac{\omega_{log}}{1.20} exp[-\frac{1.04(1+\lambda)}{\lambda-\mu^*(1+0.62\lambda)}],$$

where ω*log* is a logarithmically averaged characteristic phonon frequency, and μ$^*$ is the Coulomb pseudopotential which describes the effective electron-electron repulsion [46]. This equation is generally accurate for materials with EPC parameter λ<1.5 [47-50], whereas a numerical solution to the Eliashberg equation [42] is more adequate at larger λ. The Coulomb pseudopotential μ$^*$ is often treated as an adjustable parameter with values within a narrow range around 0.1 for most materials, making this formulism highly robust [45-49], and compares well with the latest *ab initio* Eliashberg theory [50]. Smaller values of μ$^*$ also occur, like μ$^*$=0.06 for a high-pressure phase of Si [21]. We choose 0.05 as a lower bound and 0.10 as a typical value for μ$^*$ in this study. This ranged choice for μ$^*$ is especially suitable here since the metallicity of the deformed semiconductors develops with rising strain, producing improved screening thus reduced μ$^*$, favoring the high end of predicted $T_c$ values.

## Results and Discussion

**Stress responses and metallization of shear strained Si.** To explore emerging metallic and superconducting states in deformed Si crystal, we have examined its structural and stress responses under various shear strains, which are highly effective deformation modes in generating large-range creeplike flat stress-strain curves that are indicative of metallization of covalent crystals under large strains [51]. We have identified multiple shear strain paths showing flattened stress curves with promising tendency for metallization (Supplementary Fig. 1). Here we focus primarily on the (111)[11-2] pure shear (PS) mode as an exemplary case with a pronounced creeplike stress response. Results in Fig. 1a show that stress initially rises quickly with increasing shear strain to $\varepsilon \approx 0.13$, where deformed Si becomes metallic (see below), then the rate of stress increase tapers off considerably as the degree of metallicity progressively develops, and the curve becomes nearly flat at $\varepsilon \approx 0.25$ and remains so over an extended range of further rising strain up to the dynamic stability limit $\varepsilon \approx 0.57$, beyond which imaginary phonon modes develop. This extraordinarily extended range of elastic deformation opens up vast untapped material configurations for exploring novel phenomena like reversibly tunable metallicity and superconductivity in Si, and this approach may prove generally effective for producing similar phenomena in other semiconductors.

The electronic properties of Si are highly sensitive to changes in its bonding environment. Metallic states appear in Si upon a large variety of small to moderate deformations once the Si crystal deviates from its native cubic diamond phase well before reaching the *β*-tin phase [52, 53]. Calculated results show that the electronic bandgap $E_g$ of Si drops quickly along the (111)[11-2] PS strain path, leading to a full closure at $\varepsilon=0.13$ (Fig. 1a). The electronic density of states (EDOS) at the Fermi level, $N(E_F)$, rises steadily with increasing strain, enhancing the metallicity of progressively deformed Si under various loading conditions (see Table I).

The concurrence of significant lattice softening, as indicated by the flattening stress curve, and emerging metallicity, as signified by the bandgap closure (Supplementary Fig. 2) and rising $N(E_F)$ under the (111)[11-2] shear strains in Si is dominated by an unusually large angular expansion mode with small bond elongations (Fig. 1b, c), thereby sustaining significant redistribution of bonding charge without bond breaking, reminiscent of the situation in CS deformed diamond [51]. However, in sharp contrast to the volume expanding tendency in diamond and related strong covalent solids [54-58], Si crystal undergoes a monotonic volume contraction up to dynamic stability limit at $\varepsilon=0.57$ (Fig. 1d), which stems from a large reduction of the distance between the (001) planes under the large angular deformation mode in the Si crystal (Supplementary Fig. 3). We find fairly uniform bonding charge density on both the original undeformed bonds and elongated and partially charge depleted bonds at large strains (Fig. 1e). Emerging conducting charges stemming from the depleted bonding charges are assessed by integrating the electronic states over a small (0.3 eV) energy window below the Fermi level; the results (Fig. 1f) show that these conducting charges form a well separated layered pattern as superconductivity emerges (at $\varepsilon=0.28$), but a three-dimensional pattern develops at larger strains (e.g., at ε=0.57), suggesting a crossover in transport behaviors in increasingly strained and metallic Si crystal.

**Superconductivity in strained Si.** Figure 2 shows phonon dispersion and spectral function $\alpha^2F(\omega)$ and EPC parameter $\lambda(\omega) = 2\int_0^\omega \frac{\alpha^2 F(\omega)}{\omega} d\omega$ at selected (111)[11-2] PS strains. At ε=0.28 where superconducting states start to emerge, EPC is contributed by a broad range of low- and high-frequency phonon modes, but at large strains (e.g., ε=0.57) EPC is dominated by greatly softened low-frequency modes. The nearly uniform distribution of EPC throughout the Brillouin zone is in stark contrast to CS deformed diamond where EPC is highly anisotropic and concentrated at the lattice vibrations associated with the most charge-depletion softened bonds [16]. In shear deformed Si, bonds are more uniformly stretched (Fig. 1e) with even charge distribution, leading to pronounced and nearly uniform phonon softening and EPC enhancement. It is noted that the phonon softening is intimately related to the flattening of the stress curves (Supplementary Fig. 4), both stemming from the strain weakened bonding structures. Also shown in Fig. 2 are the electronic band structures at the selected shear strains, and it is seen that increasing strain leads to more extensive band crossing at the Fermi level, which would significantly impact electronic transport behaviors.

To test robustness of superconductivity in shear strained Si, we have examined results under the (11-2)[111] CS and (111)[11-2] tension shear (TS) (Supplementary Fig. 5). The results in Fig. 3 (see Supplementary Fig. 6 for electronic band structures at the selected strains) show the same patterns of flattened stress curves, quick bandgap reduction and closure, and ensuing emergence of superconducting states, although the rise of $T_c$ is limited by shortened dynamic stability range compared to the PS case. There is clear dominance by low-frequency modes on the phonon softening and EPC enhancement, demonstrating that the same robust mechanism is driving superconductivity in all these differently deformed Si lattices. These results establish consistent trends in diverse shear-strain enhanced $N(E_F)$, softened $\omega_{\log}$, and higher $\lambda$ and $T_c$ (see Table I).

**Metallization and superconductivity in strained SiC.** We now turn to SiC, which comprises a C-Si bonding network in diamond structure and a much larger band gap compared to Si (Supplementary Fig. 2). We examine the (111)[11-2] PS that has the most flattened stress curve among the examined cases (Supplementary Fig. 1c). Compared to Si, the bandgap of SiC decreases with rising strain at a slightly reduced but still fast rate, and the metallic then superconducting states develop quickly as indicated in Fig. 4a, which reinforces the notion that Si and SiC share key structural and electronic response patterns under shear loading conditions, raising excellent prospects that these behaviors are commonly present in shear-strained semiconductors. Interestingly, calculated results in Fig. 4b show that EPC in shear deformed SiC is mainly contributed by phonon modes across a broad frequency range but concentrated in momentum space near the Γ and A points, which is similar to the situation in CS deformed diamond [16] but different from the nearly uniform distributions in Si under diverse PS, CS, or TS strains (see results in Fig. 2a and Fig. 3b, e). Despite such different EPC characteristics, Si and SiC share the most prominent features of shear-strain induced robust metallization and strong phonon softening, leading to substantial superconducting states in distinct material environments. Such diversity bodes well for finding similar phenomena driven by the same general mechanisms in many other semiconductors.

Carrier doping has produced superconductivity with $T_c$ of 0.35 K in Si [22] and 1.4 K in SiC [26]. Here, we show that diverse shear strains can effectively modulate $T_c$ over extended ranges up to much higher values of 8.1 K for Si and 3.7 K for SiC. These results underscore excellent prospects for inducing vigorous superconducting states in elastically deformed semiconductors. These superior $T_c$ values under tunable shear strains are well above those achieved in semiconducting or conducting films under fixed tensile strains by lattice mismatch [11-14], making shear-strain tuning a desirable approach.

High pressure can effectively modulate material properties [59-61], but stringent equipment requirements impose a very high barrier for most practical implementation; in contrast, strain tuning is compatible with advanced material characterization protocols, and avoids the need for external carriers, usually by ion implantation that commonly causes structural and charge disorder with uncontrolled complications in the host material [1, 8]. Tuning superconductivity in elastically deformed semiconductors allows for substantial property modulations, as seen in a recent experiment showing that soft phonon modes associated with unique bond contractions induced by interfacial strains induced surprising superconductivity in $RuO_2$ films [14]. Such coherent epitaxy is the current state of the art in strain engineering [11-14]. The latest nanoscale fabrication and imaging techniques have demonstrated reversible ultralarge elastic deformation approaching the theoretical limits under diverse loading conditions [62-66]. These powerful techniques open up vast structural configuration space for exploring richer material behaviors and physics processes beyond the traditional paradigm of structural stability at quasi-hydrostatic pressures to discover novel material behaviors and enable new functionalities with major technological impact.

**Conclusions**

In summary, we have uncovered via first-principles calculations robust superconducting states in shear strained Si and SiC crystals under diverse loading conditions. Analysis of structural, electronic, phonon, and electron-phonon coupling behaviors under evolving strains reveals consistent patterns and trends of shear deformation induced and enhanced metallization and phonon softening, resulting in vigorous superconducting states that host rising $T_c$ at progressively increasing strains. This work opens a promising avenue for inducing and tuning superconducting states in semiconductors by versatile shear strains. The key physics processes and mechanisms unveiled here are insensitive to material details and, therefore, are expected to be general phenomena not limited to Si and SiC or cubic crystals, and should remain robust and applicable to wide-ranging semiconductors. While covalent crystals provide a favorable platform for the physics described here, because of their highly concentrated and directional bonding structures, we expect that strain induced charge redistribution and phonon softening provide an effective approach for tuning superconductivity in a broad variety of materials. Advanced nanoscale synthesis and characterization techniques have produced high-quality specimens and tested mechanical properties [62-67], and these techniques are also well equipped to operate at cryogenic conditions for versatile physical property measurements at extremely low temperatures [68-70], making experimental verification of our predicted results feasible.

The present findings have major implications beyond tuning superconducting states in semiconductors, e.g., for deciphering exceptional phonon-mediated superconductors like the long-sought metallic hydrogen [71, 72] and recently discovered superhydrides [73-76], which have been predicted or demonstrated to superconduct near room temperature at megabar pressures. Non-hydrostatic conditions like uniaxial compression and shear strains are inevitably present in such extreme environments, exerting major influence on material behaviors, such as incipient structural instability that has been postulated to impact superconductivity in $LaH_{10}$ [77].

Current experimental methods employing traditional or rotational diamond anvil cell setups are capable of producing large shear-strain conditions at high pressures, and recent studies using these setups have revealed significant shear-induced property modulations such as altered kinetics of pressure induced graphite-to-diamond transition [78] and discovery of a new shear-driven formation route for diamond [79]. Such capabilities provide a powerful technique for probing shear strain effects on diverse phonon-mediated superconductors, e.g., high-$T_c$ superhydrides. This study provides a strong impetus for exploring effects of broadly defined strains on modulating these fascinating superconducting materials. Work along this line may prove fruitful in discovering intriguing phenomena and elucidating key mechanisms.

**Supplementary material**

Supplementary material is available online

**Funding**

This work was mainly supported by the National Key Research and Development Program of China (Grant No. 2018YFA0703400), Natural Science Foundation of China (Grant Nos. 12074140 and 12034009), China Postdoctoral Science Foundation (Grant No. 2020M681031), and Program for JLU Science and Technology Innovative Research Team (JLUSTIRT). Reported calculations utilized computing facilities at the High-Performance Computing Center of Jilin University and Tianhe2-JK at the Beijing Computational Science Research Center.

# References

[1] Blase X, Bustarret E, Chapelier C, Klein T and Marcenat C 2009 *Nat. Mater.* 8 375

[2] Gurevich V L, Larkin A I and Firsov Y A 1962 *Sov. Phys. Solid State* 4 131

[3] Cohen M L 1964 *Phys. Rev.* 134 A511

[4] Cohen M L 1964 *Rev. Mod. Phys.* 36 240

[5] Schooley J F, Hosler W R and Cohen M L 1964 *Phys. Rev. Lett.* 12 474

[6] Schooley J F, Hosler W R, Ambler E, Becker J H, Cohen M L and Koonce C S 1965 *Phys. Rev. Lett.* 14 305

[7] Hein R A, Gibson J W, Mazelsky R, Miller R C and Hulm J K 1964 *Phys. Rev. Lett.* 12 320

[8] Bustarret E 2015 *Physica C* 514 36

[9] Mao H K, Chen X J, Ding Y, Li B and Wang L 2018 *Rev. Mod. Phys.* 90 015007

[10] Liu Z, Dong Q, Shan P, Wang Y, Dai J, Jana R, Chen K, Sun J, Wang B, Yu X, Liu G, Uwatoko Y, Sui Y, Yang H, Chen G and Cheng J 2020 *Chin. Phys. Lett.* 37 047102

[11] Hicks C W, Brodsky D O, Yelland E A, Gibbs A S, Bruin J A N, Barber M E, Edkins S D, Nishimura K, Yonezawa S, Maeno Y, Mackenzie A P 2014 *Science* 344 283

[12] Ahadi K, Galletti L, Li Y, Salmani-Rezaie S, Wu W and Stemmer S 2019 *Sci. Adv.* 5 eaaw0120

[13] Wu X, Ming F, Smith T S, Liu G, Ye F, Wang K, Johnston S and Weitering H H 2020 *Phys. Rev. Lett.* 125 117001

[14] Uchida M, Nomoto T, Musashi M, Arita R and Kawasaki M 2020 *Phys. Rev. Lett.* 125 147001

[15] Yuan Y H, Wang X T, Song C L, Wang L L, He K, Ma X C, Yao H, Li W and Xue Q K 2020 *Chin. Phys. Lett.* 37 017402

[16] Liu C, Song X, Li Q, Ma Y M and Chen C F 2020 *Phys. Rev. Lett.* 124 147001

[17] Chang K J and Cohen M L 1984 *Phys. Rev. B* 30 5376

[18] Olijnyk H, Sikka S K and Holzapfel W B 1984 *Phys. Lett. A* 103 137

[19] Hu J Z and Spain I L 1984 *Solid State Commun.* 51 263

[20] Mignot J M, Chouteau G and Martinez G 1985 *Physica B* 135 235

[21] Chang K J, Dacorogna M M, Cohen M L, Mignot J M, Chouteau G and Martinez G 1985 *Phys. Rev. Lett.* 54 2375

[22] Bustarret E, Marcenat C, Achatz P, Kacmarcik J, Levy F, Huxley A, Ortega L, Bourgeois E, Blase X, Debarre D and Boulmer J 2006 Nature 444 465

[23] Ekimov E A, Sidorov V A, Bauer E D, Melnik N N, Curro N J, Thompson J D and Stishov S M 2004 Nature 428 542

[24] Takano Y, Nagao M, Sakaguchi I, Tachiki M, Hatano T, Kobayashi K, Umezawa H and Kawarada H 2004 Appl. Phys. Lett. 85 2851

[25] Zhang G, Turner S, Ekimov E A, Vanacken J, Timmermans M, Samuely T, Sidorov V A, Stishov S M, Lu Y, Deloof B, Goderis B, Van Tendeloo G, Van de Vondel J and Moshchalkov V V 2014 Adv. Mater. 26 2034

[26] Ren Z A, Kato J, Muranaka T, Akimitsub J, Kriener M and Maeno Y 2007 J. Phys. Soc. Jpn. 76 103710

[27] Kresse G and Furthmüller L 1996 Phys. Rev. B 54 11169

[28] Perdew J P and Zunger A 1981 Phys. Rev. B 23 5048
[29] Ceperley D M and Alder B J 1980 Phys. Rev. Lett. 45 566
[30] Giannozzi P et al. 2009 J. Phys.: Condens. Matter 21 395502
[31] Boeri L, Kortus J and Andersen O K 2004 Phys. Rev. Lett. 93 237002
[32] Lee K W and Pickett W E 2004 Phys. Rev. Lett. 93 237003
[33] Xiang H J, Li Z Y, Yang J L, Hou J G and Zhu Q S 2004 Phys. Rev. B 70 212504
[34] Blase X, Adessi C and Connetable D 2004 Phys. Rev. Lett. 93 237004
[35] Ma Y, Tse J S, Cui T, Klug D D, Zhang L, Xie Y, Niu Y and Zou G 2005 Phys. Rev. B 72 014306
[36] Giustino F, Yates J R, Souza I, Cohen M L and Louie S G 2007 Phys. Rev. Lett. 98 047005
[37] Noffsinger J, Giustino F, Louie S G and Cohen M L 2009 Phys. Rev. B 79 104511
[38] Heyd J, Scuseria G E and Ernzerhof M 2003 J. Chem. Phys. 118 8207
[39] Heyd J, Scuseria G E and Ernzerhof M 2006 Erratum: J. Chem. Phys. 124 219906
[40] Bludau W, Onton A and Heinke W 1974 J. Appl. Phys. 45 1846
[41] Hunphreys R G, Bimberg D and Choyke W J 1981 Solid State Commun. 39 163
[42] Eliashberg G M 1960 Sov. Phys. JETP 11 696
[43] Scalapino D J, Schrieffer J R and Wilkins J W 1966 Phys. Rev. 148 263
[44] McMillan W L 1968 Phys. Rev. 167 331
[45] Allen P B and Dynes R C 1975 Phys. Rev. B 12 905
[46] Morel P and Anderson P W 1962 Phys. Rev. 125 1263
[47] McMillan W L and Rowell J M 1969 Superconductivity edited by Parks R D (New York: Marcel Dekker) vol 1 p561
[48] Carbotte J P 1990 Rev. Mod. Phys. 62 1027
[49] Carbotte J P and Marsiglio F 2003 Electron-Phonon Superconductivity in The Physics of Superconductors edited by Bennemann K H and Ketterson J B (Berlin: Heidelberg)
[50] Sanna A, Flores-Livas J A, Davydov A, Profeta G, Dewhurst K, Sharma1 S and Gross E K U 2018 J. Phys. Soc. Japn. 87 041012
[51] Liu C, Song X, Li Q, Ma Y M and Chen C F 2019 Phys. Rev. Lett. 123 195504
[52] Shen G, Ikuta D, Sinogeikin S, Li Q, Zhang Y and Chen CF 2012 Phys. Rev. Lett. 109 205503
[53] Zarkevich N A, Chen H, Levitas V I and Johnson D D 2018 Phys. Rev. Lett. 121 165701
[54] Zhang Y, Sun H and Chen C F 2005 Phys. Rev. Lett. 94 145505
[55] Zhang Y, Sun H and Chen C F 2006 Phys. Rev. B 73 144115
[56] Zhang Y, Sun H and Chen C F 2006 Phys. Rev. B 73 064109
[57] Li B, Sun H and Chen C F 2014 Nat. Commun. 5 4965
[58] Li B, Sun H and Chen C F 2016 Phys. Rev. Lett. 117 116103
[59] Chen X, Zhan X H, Wang X J, Deng J, Liu X B, Chen X, Guo J G and Chen X L 2021 Chin. Phys. Lett. 38 057402
[60] Gu Q Y, Xing D Y and Sun J 2019 Chin. Phys. Lett. 36 097401
[61] Zhang X, Luo T C, Hu X Y, Guo J, Lin G C, Li Y H, Liu Y Z, Li X K, Ge J, Xing Y, Zhu Z W, Gao P, Sun L L and Wang J 2019 Chin. Phys. Lett. 36 057402
[62] Zhang H, Tersoff J, Chen S X H, Zhang Q, Zhang K, Yang Y, Lee C S, Tu K N, Li J and Lu Y 2016 Sci. Adv. 2 e1501382

[63] Banerjee A, Bernoulli D, Yuen H Z M F, Liu J, Dong J, Ding F, Lu J, Dao M, Zhang W, Lu Y and Suresh S 2018 Science 360 300

[64] Nie A, Bu Y, Li P, Zhang Y, Jin T, Liu J, Su Z, Wang Y, He J, Liu Z, Wang H, Tian Y and Yang W 2019 Nat. Commun. 10 5533

[65] Chen M, Pethö L, Sologubenko A S, Ma H, Michler J, Spolenak R and Wheeler J M 2020 Nat. Commun. 11 2681

[66] Shi Z, Dao M, Tsymbalov E, Shapeev A, Li J and Suresh S 2020 Proc. Natl. Acad. Sci. USA 117 24634

[67] Dang C, Chou J, Dai B, Chou C, Yang Y, Fan R, Lin W, Meng F, Hu A, Zhu J, Han J, Minor A M, Li J and Lu Y 2021 Science 371 76

[68] Luo W, Boselli M, Poumirol J, Ardizzone I, Teyssier J, van der Marel D, Gariglio S, Triscone J and Kuzmenko A B 2019 Nat. Commun. 10 2774

[69] Jiang S, Xie H, Shan J and Mak K F 2020 Nat. Mater. 19 1295

[70] Tang Y, Li L, Li T, Xu Y, Liu S, Barmak K, Watanabe K, Taniguchi T, MacDonald A H, Shan J and Mak K F 2020 Nature 579 353

[71] Ashcroft N W 1968 Phys. Rev. Lett. 21 1748

[72] McMahon J M and Ceperley D M 2011 Phys. Rev. B 84 144515

[73] Sun Y, Lv J, Xie Y, Liu H and Ma Y M 2019 Phys. Rev. Lett. 123 097001

[74] Drozdov A P, Kong P P, Minkov V S, Besedin S P, Kuzovnikov M A, Mozaffari S, Balicas L, Balakirev F F, Graf D E, Prakapenka V B, Greenberg E, Knyazev D A, Tkacz M and Eremets M I 2019 Nature 569 528

[75] Somayazulu M, Ahart M, Mishra A K, Geballe Z M, Baldini M, Meng Y, Struzhkin V V and Hemley R J 2019 Phys. Rev. Lett. 122 027001

[76] Hong F, Yang L X, Shan P F, Yang P T, Liu Z Y, Sun J P, Yin Y Y, Yu X H, Cheng J G and Zhao Z X 2020 Chin. Phys. Lett. 37 107401

[77] Sun D, Minkov V S, Mozaffari S, Chariton S, Prakapenka V B, Eremets M I, Balicas L and Balakirev F F 2020 arXiv:2010.00160 [cond-mat.supr-con]

[78] Gao Y, Ma Y Z, An Q, Levitas V, Zhang Y, Feng B, Chaudhuri J and Goddard W A 2019 Carbon 146 364

[79] Dong J, Yao Z, Yao M, Li R, Hu K, Zhu L, Wang Y, Sun H, Sundqvist B, Yang K and Liu B 2020 Phys. Rev. Lett. 124 065701

**Figure Captions**

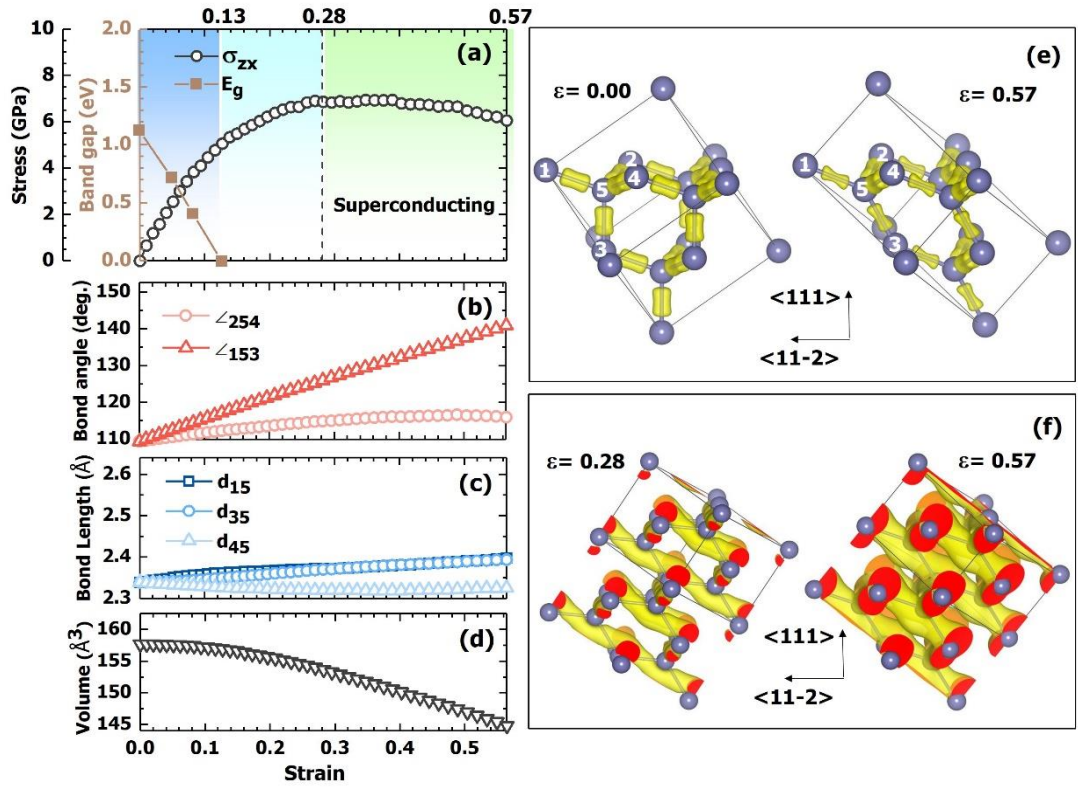

**Fig. 1.** (a) Stress and bandgap evolution of Si under the (111)[11-2] PS strain up to the dynamic stability limit ε=0.57. Metallization and superconducting states emerge at ε>0.13 and ε>0.28, respectively. Variations of (b) bond angle $α_{153}$ between atom pairs 1-5 and 5-3 and $α_{254}$ [see panel (e)], (c) three key bond lengths, $d_{15}$, $d_{35}$ and $d_{45}$, and (d) the unitcell volume. (e,f) Structural snapshots with bonding charge density at selected strains.

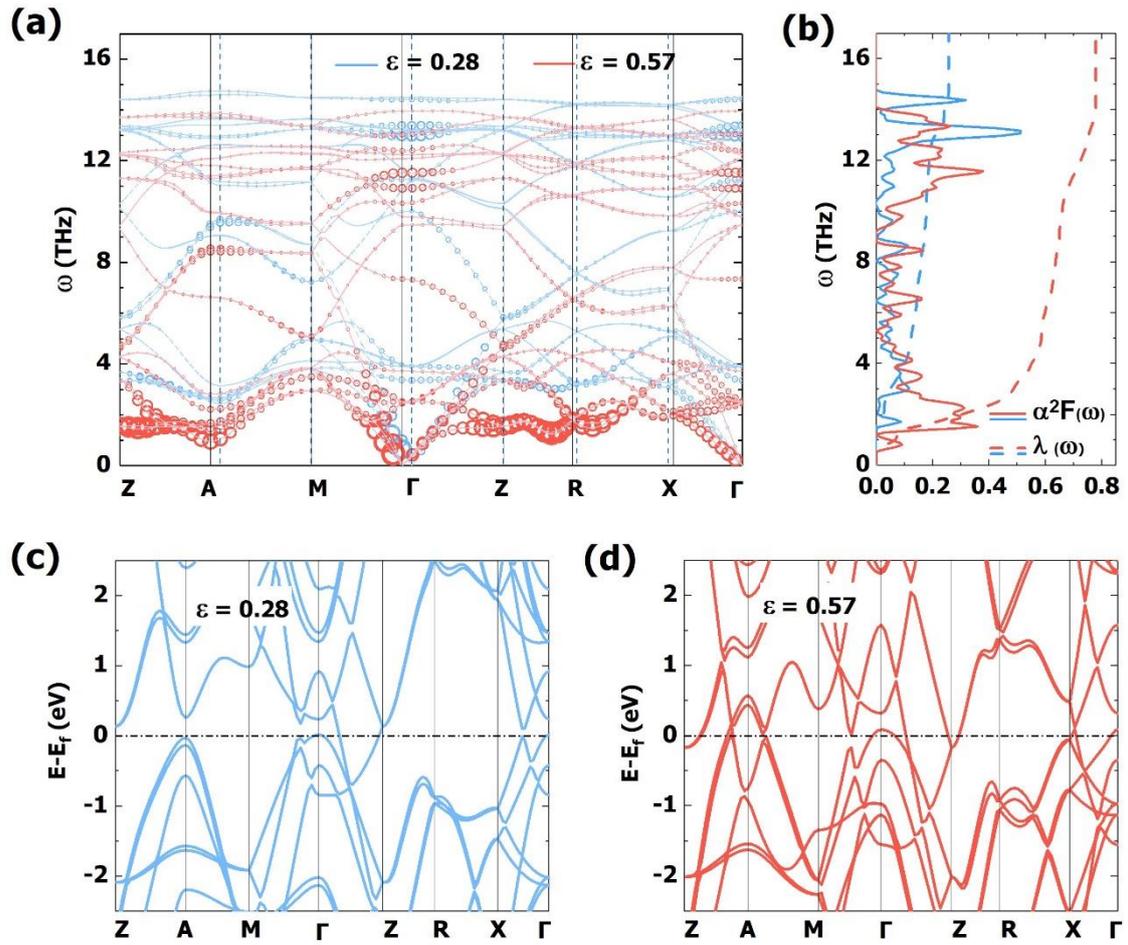

**Fig. 2.** (a) Phonon dispersions at selected (111)[11-2] PS strains with the strength of q-resolved $\lambda_q$ indicated by circle size. The solid and dashed vertical lines indicate (also in Fig. 3b, e and 4b) the positions of the high-symmetry points in the reciprocal lattice at different indicated shear strains. (b) Spectral function $\alpha^2F(\omega)$ and $\lambda(\omega)$. (c,d) Calculated electronic band structures at ε=0.28. and ε=0.57, respectively.

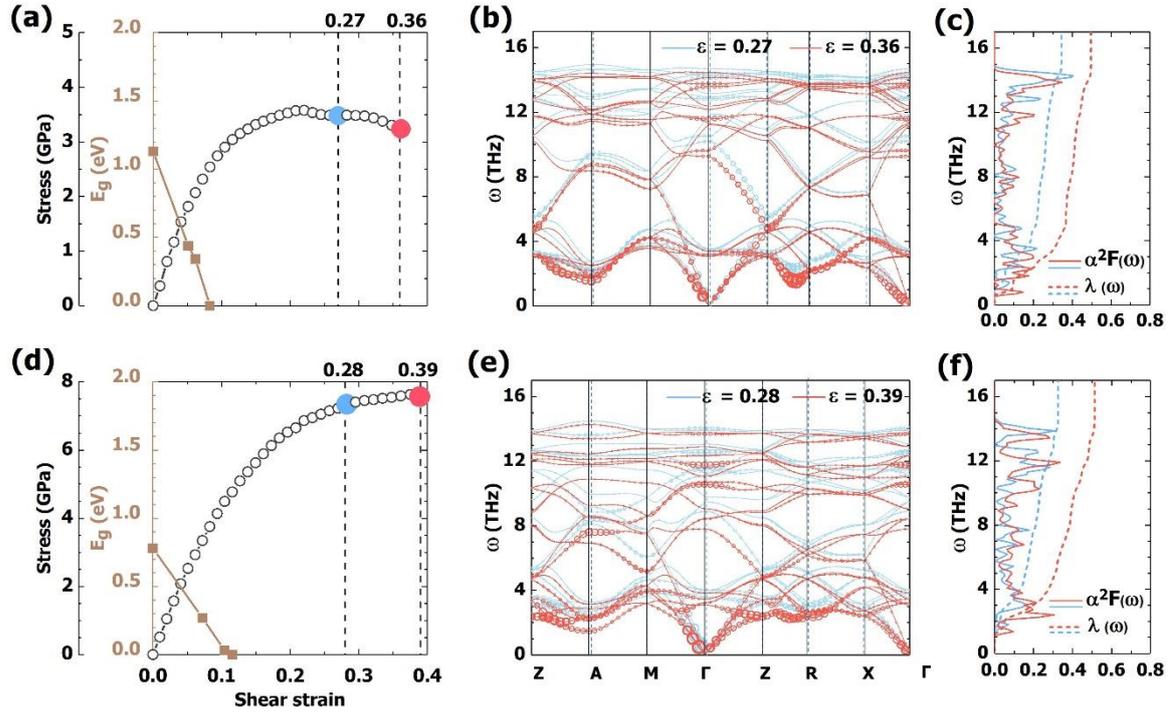

**Fig. 3.** (a,d) Stress and bandgap evolution of Si under (11-2)[111] CS or (111)[11-2] TS strains. (b,e) Phonon dispersions at selected strains with the strength of q-resolved $\lambda_q$ indicated by circle size. (c,f) Spectral function $\alpha^2F(\omega)$ and $\lambda(\omega)$.

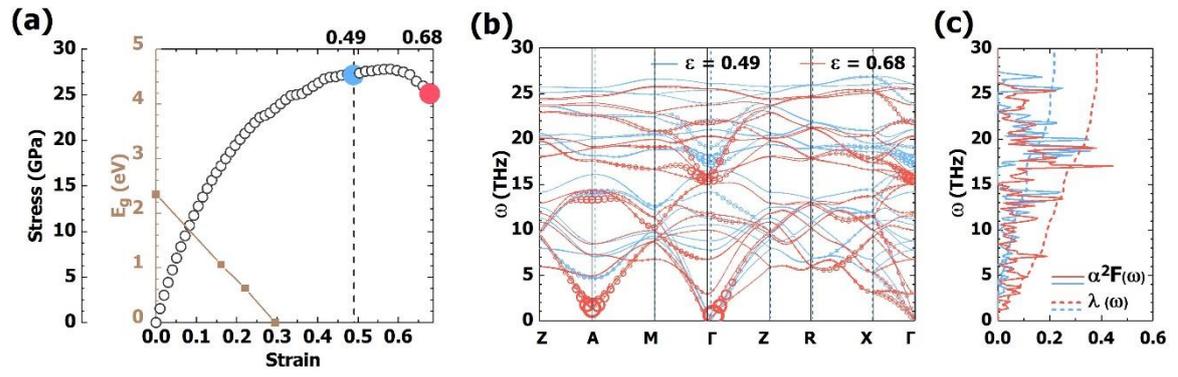

**Fig. 4.** (a) Stress and bandgap evolution of SiC under (111)[11-2] PS strains. (b) Phonon dispersions at selected strains with the strength of q-resolved $\lambda_q$ indicated by circle size. (c) Spectral function $\alpha^2 F(\omega)$ and $\lambda(\omega)$.

**Table 1.** Computed $N(E_F)$ (states/spin/Ry/unitcell), $\omega_{log}(K)$, $\lambda$ and $T_c$ (K) for Si and SiC at various PS, CS or TS strains. Two $T_c$ values set the range for $\mu^* = 0.10$ and 0.05.

| Loading path | Strain | $N(E_F)$ | $\omega_{log}$ | $\lambda$ | $T_c$ |
| --- | --- | --- | --- | --- | --- |
| Si PS (111)[11-2] | 0.28 | 9.7 | 282 | 0.26 | 0.0-0.4 |
| Si PS (111)[11-2] | 0.42 | 13.3 | 219 | 0.42 | 1.3-2.9 |
| Si PS (111)[11-2] | 0.53 | 15.8 | 157 | 0.67 | 4.8-7.2 |
| Si PS (111)[11-2] | 0.57 | 16.6 | 145 | 0.78 | 6.5-8.7 |
| Si CS (11-2)[111] | 0.22 | 9.5 | 260 | 0.26 | 0.0-0.3 |
| Si CS (11-2)[111] | 0.27 | 11.0 | 218 | 0.35 | 0.3-1.4 |
| Si CS (11-2)[111] | 0.36 | 12.3 | 153 | 0.50 | 1.8-3.5 |
| Si TS (111)[11-2] | 0.28 | 11.0 | 261 | 0.33 | 0.3-1.2 |
| Si TS (111)[11-2] | 0.39 | 14.6 | 207 | 0.51 | 2.8-5.1 |
| SiC PS (111)[11-2] | 0.49 | 5.3 | 628 | 0.22 | 0.0-0.2 |
| SiC PS (111)[11-2] | 0.68 | 6.4 | 393 | 0.38 | 1.3-3.7 |